\documentclass[paper]{JHEP3}
\usepackage{epsfig}
\usepackage{graphicx}
\usepackage{xspace}
\usepackage{amssymb}
\usepackage{subfigure}
\usepackage{multirow}
\newcommand{\ra}{\rightarrow}

\newcommand{\be}{\begin{equation}}
\newcommand{\ee}{\end{equation}}

\newcommand{\bea}{\begin{eqnarray}}
\newcommand{\eea}{\end{eqnarray}}
\newcommand{\beanon}{\begin{eqnarray*}}
\newcommand{\eeanon}{\end{eqnarray*}}
\newcommand{\ba}{\begin{array}}
\newcommand{\ea}{\end{array}}
\newcommand{\bd}{\begin{description}}
\newcommand{\ed}{\end{description}}
\newcommand{\bi}{\begin{itemize}}
\newcommand{\ei}{\end{itemize}}
\newcommand{\ben}{\begin{enumerate}}
\newcommand{\een}{\end{enumerate}}
\newcommand{\bc}{\begin{center}}
\newcommand{\ec}{\end{center}}

\newcommand{\eqn}[1]{Eq.(\ref{#1})}

\newcommand{\tbn}[1]{Tab.~\ref{#1}}
\newcommand{\tbns}[2]{Tabs.~\ref{#1}--\ref{#2}}

\newcommand{\fig}[1]{Fig.~\ref{#1}}

\newcommand{\sect}[1]{Sect.~\ref{#1}}

\newcommand{\rf}[1]{Ref.~\cite{#1}}
\newcommand{\rfs}[1]{Refs.~\cite{#1}}  % comma separated list

\newcommand{\Phantom}{{\tt PHANTOM}\xspace}

\newcommand{\MadEvent}{{\tt MADEVENT}\xspace}

\def\pl #1 #2 #3 {{\it Phys.~Lett.} {\bf#1} (#2) #3}   
\def\np #1 #2 #3 {{\it Nucl.~Phys.} {\bf#1} (#2) #3}
\def\zp #1 #2 #3 {{\it Z.~Phys.} {\bf#1} (#2) #3}
\def\pr #1 #2 #3 {{\it Phys.~Rev.} {\bf#1} (#2) #3}
\def\prep #1 #2 #3 {{\it Phys.~Rep.} {\bf#1} (#2) #3}
\def\prl #1 #2 #3 {{\it Phys.~Rev.~Lett.} {\bf#1} (#2) #3}
\def\intj #1 #2 #3 {{\it Int. J. Mod. Phys.} {\bf#1} (#2) #3}
\def\mpl #1 #2 #3 {{\it Mod.~Phys.~Lett.} {\bf#1} (#2) #3}
\def\rmp #1 #2 #3 {{\it Rev. Mod. Phys.} {\bf#1} (#2) #3}
\def\cpc #1 #2 #3 {{\it Comp. Phys. Commun.} {\bf#1} (#2) #3}
\def\epj #1 #2 #3 {{\it Eur. Phys. J.} {\bf#1} (#2) #3}
\def\jhep #1 #2 #3 {{\it JHEP} {\bf#1} (#2) #3}

\title{Multiple Parton Interactions in $Z+jets$ production at the LHC.
A comparison of factorized and non--factorized double parton distribution functions.
}

\author{
Ezio Maina$^{a,b}$\\
$^a$ INFN, Sezione di Torino, Italy,\\
$^b$ Dipartimento di Fisica Teorica, Universit\`a di Torino, Italy
}

\preprint{DFTT 21/2010}

\abstract{

We examine the contribution of Multiple Parton Interactions to $Z+n$--jets production at
the LHC, $n=2,3,4$, where the $Z$ boson is assumed to decay leptonically.

We compare the results obtained with the correlated GS09 double parton
distribution function with those obtained with two instances
of fully factorized single parton distribution functions: MSTW2008LO and CTEQ6L1.

It appears quite feasible to measure the MPI contribution to  
$Z$+2/3/4 jets already in the first phase of the LHC with a total
luminosity of one inverse femtobarn at 7 TeV. If as expected the trigger threshold
for single photons is around $80~{\rm GeV}$, $Z+2$--jets production
may well turn out to be more easily observable than the $\gamma+3$--jets channel.
The MPI cross section is dominated by relatively soft events with two jets
balancing in transverse momentum.

}  

\begin{document}

%%%%%%%%%%%%%%%%%%%%%%%%%%%%%%%%%%%%%%%%%%%%%%%%%%%%%%%%%%%%%%%%%%%%%%%%%%%%%%%%

\section{Introduction}
\label{sec:intro}

The QCD--improved Parton Model forms the basis of our understanding of high--energy
hadron scattering. In this framework each hadron is described as a collection of essentially free
elementary constituents. The interactions between constituents belonging to different colliding hadrons
are the seeds of the complicated process which eventually leads to the particles observed in the detector.
In this conceptual scheme it is quite natural to envisage the possibility that more than one pair
of partons might interact in a single hadronic impact. This kind of events are referred to as
Multiple Parton Interactions (MPI) while those in which only a single pair of partons
produce a hard scattering are described as Single Parton Interactions (SPI).

Multiple Parton Interactions have been detected in high transverse momentum hadron
collisions both at the ISR at CERN \cite{Akesson:1986iv} and at the Tevatron at Fermilab  
\cite{Abe:1997bp,Abe:1997xk,Abazov:2009gc}. The measured cross sections imply that MPI
could provide a non-negligible background to all sort of interesting reactions
since MPI rates at the LHC are expected to be large.
At smaller transverse momentum MPI have been shown to be necessary for the 
successful description of the underlying event both in Pythia
\cite{Sjostrand:1987su,Sjostrand:2004pf,Sjostrand:2004ef} and in Herwig
\cite{Butterworth:1996zw,Bahr:2008dy}. The wide range of phenomena in which MPI
are involved highlights the urgency of a more thorough understanding of these
reactions both experimentally and from a theoretical point of view. 

The theoretical investigation of MPI has a long history
\cite{Landshoff:1978fq,Takagi:1979wn,Goebel:1979mi,Paver:1982yp,Mekhfi:1985dv}
and have experienced a renewed interest in more recent times
\cite{Snigirev:2003cq,Korotkikh:2004bz,Treleani:2007gi,Calucci:2008jw,Calucci:2009sv,
Calucci:2009ea,Snigirev:2010tk,Calucci:2010wg,Snigirev:2010ds}.

The basic formalism can be readily described starting from the
standard expression for the SPI cross section:
\be
\label{eq:sigma_S1}
\sigma^S_{(A)} = \sum_{i,j}\int F_i(x_1,t_1)\sigma^A_{ij}(x_1,y_1)F_j(y_1,t_1)\,dx_1 dy_1
\ee
where $t_1$ is the factorization scale which characterizes the interaction
and at which
the parton distribution functions $F_i(x_1,t_1)$ for parton $i$ to have momentum
fraction $x_1$ are evaluated.
\eqn{eq:sigma_S1} can be rewritten in term of parton distributions
which depend on the transverse
coordinates as well as on the longitudinal momentum fraction as:
\be
\label{eq:sigma_S2}
\sigma^S_{(A)} = 
        \sum_{i,j} \int\Gamma_i (x_1,b_1,t_1)\sigma^A_{ij}(x_1,y_1)\Gamma_j (y_1,b_1 - \beta,t_1)
        \, dx_1\, dy_1 \, d^2 b_1\, d^2\beta
\ee
where $\beta$ is the usual impact parameter.
Making the reasonable assumption that the dependence on the momentum fraction 
and that on the transverse position factorize 
\be
\label{eq:Gamma_fact}
\Gamma_i (x,b) = F_i(x) \times f(b)
\ee
and that the latter is a universal function for all
kind of partons fixes the normalization of the transverse distribution:
\be
\label{eq:overlap}
\int f(b)f(b - \beta) \, d^2 b \, d^2\beta = \int  T(\beta)\, d^2\beta = 1
\ee
where we have defined the overlap function  $T(\beta) = \int f(b)f(b - \beta) \, d^2 b$.

Analogously we can write the Double Parton Interaction (DPI)
cross section as follows:

\bea
\label{eq:sigma_D1}
\sigma^D_{(A,B)} &=& \frac{m}{2!} \sum_{i,j,k,l} \int \Gamma_{i,j} (x_1,b_1,t_1,x_2,b_2,t_2)
\sigma^A_{i,k}(x_1,y_1)\, \sigma^B_{j,l}(x_2,y_2) \\[-4mm] 
         & & \hspace{2.1cm} \times \, \Gamma_{k,l} (y_1,b_1 - \beta,t_1,y_2,b_2 - \beta,t_2) 
         \, dx_1\, dy_1 \,d^2 b_1dx_2 \, dy_2 \, d^2 b_2 \, d^2\beta \nonumber
\eea
where $t_1$ and $t_2$ are the factorization scales of the two scatterings;
$m$ is a symmetry factor which is equal to one if the reactions $A$ and $B$ are
identical and equal to two if they are not.

Separating the transverse part,
$\Gamma_{i,j} (x_1,b_1,t_1,x_2,b_2,t_2)= F_{i,j} (x_1,t_1,x_2,t_2)\times f(b_1)\times f(b_2)$
\eqn{eq:sigma_D1} becomes
\bea
\label{eq:sigma_D2}
\sigma^D_{(A,B)} &=& \frac{m}{2!\,\sigma_{eff}} \sum_{i,j,k,l} \int F_{i,j} (x_1,t_1,x_2,t_2)
\sigma^A_{i,k}(x_1,y_1)\, \sigma^B_{j,l}(x_2,y_2) \\[-4mm] 
         & & \hspace{4.5cm} \times \, F_{k,l} (y_1,t_1,y_2,t_2) 
         \, dx_1\, dy_1 dx_2 \, dy_2 \nonumber
\eea
where
\be
\label{eq:sigma2_eff}
\frac{1}{\sigma_{eff}}=  \int  T^2(\beta)\, d^2\beta.
\ee

If one makes the further assumptions that double parton distributions
reduce to the product of
two independent one parton distributions, $F_{i,j} = F_{i}\times F_{j}$,
the DPI cross section can be expressed in the simple form
\be
\label{eq:sigma_D3}
\sigma^D_{(A,B)} = \frac{m}{2!} \frac{\sigma^S_{(A)}\sigma^S_{(B)}}{\sigma_{eff}}.
\ee

This last assumption however, even though rather common in the literature and
quite convenient from a computational point of view, is clearly incorrect.
In Ref.~\cite{Snigirev:2003cq,Korotkikh:2004bz} it was shown that
correlations between the value of the double distribution functions for
different values of the two momentum fractions $x_1, x_2$ are to be expected,
even under the assumption of no correlation at some scale $Q_0$, as a
consequence of the evolution of the distribution functions 
to a different scale $Q$, which is determined
by an equation analogous to the usual DGLAP equation
\cite{Kirschner:1979im,Zinovev:1982be,Shelest:1982dg}.

A large number of studies have evaluated the MPI contribution to several high energy
processes \cite{Paver:1983hi,Humpert:1983pw,Paver:1984ux,Humpert:1984ay,
Ametller:1985tp,Halzen:1986ue,Mangano:1988sq,Godbole:1989ti,Drees:1996rw,
DelFabbro:1999tf,DelFabbro:2002pw,Hussein:2006xr,Hussein:2007gj,Domdey:2009bg},
including Higgs and electroweak vector boson production.
Other studies \cite{Kulesza:1999zh,Maina:2009vx,Maina:2009sj}
have in addition focused on the differences between final states produced
in SPI and in MPI as a tool to reduce the background due to MPI or alternatively
to separate MPI processes from SPI ones and gain more detailed experimental
information on Multiple Parton Interactions.
All the aforementioned studies have assumed complete factorization of double Parton
Distribution Functions (dPDF).

In \cite{Cattaruzza:2005nu} the corrections to the factorized form for the 
dPDF have been estimated. They depend on the
factorization scale, being larger at larger scales $Q$, and on the $x$ range,
again being more important at larger momentum fractions. For $Q = M_W$ and
$x \sim 0.1$ the corrections are about 35\% for the gluon-gluon case.    

Recently Gaunt and Stirling \cite{Gaunt:2009re} have developed a set of dPDF
which satisfy a collection of momentum and number sum rules. These sum rules are
preserved by the evolution equations \cite{Snigirev:2003cq,Korotkikh:2004bz}
and therefore are obeyed at any scale $Q$ once they are satisfied at an input scale
$Q_0$. The GS09 set is based on the MSTW2008LO
single Parton Distribution Functions (sPDF) \cite{Martin:2009iq}.
Gaunt and Stirling also provide a program which evolves the dPDF from the input scale
to any scale and a set of publicly available dPDF grids. In their published
form the GS09 set deals with the case of two identical scales $t_1$ and $t_2$
in the distribution functions  $F_{i,j}$, but this limitation has been recently
dropped.

In Ref.~\cite{Gaunt:2010pi} the GS09 dPDF have been employed in a study of
same--sign $W$ pair production at the LHC, including the background due to
$W^\pm Z(\gamma^*)$ production. From the ratio 
$R \equiv 4\,\sigma_{W^+W^+}\sigma_{W^-W^-}/\sigma^2_{W^+W^-}$ , which is equal to one
for factorized dPDF,
a violation of factorization at the 20\% to 30\% level is reported.

In this paper we examine the contribution of MPI to $Z+n$--jets production at
the LHC, $n=2,3,4$, where the $Z$ boson is assumed to decay leptonically.
These processes have the advantage of a much larger cross section than same--sign
$WW$ production and therefore are more likely to allow detailed studies of MPI
at the low luminosity, about 1 fb$^{-1}$, foreseen for the first two years of
operation at the LHC with $\sqrt{s} = 7$ TeV. While the cross section for $Z+nj$
is smaller than for $W+nj$ because of the smaller leptonic branching ratio,
the former is cleaner from an experimental point of view
since isolated, high pT charged leptons, which are the main
tool for $W$ detection, can be copiously produced in B-hadron decays
\cite{Sullivan:2006hb,Sullivan:2008ki,Sullivan:2010jk}
while no comparable mechanism exists for generating
lepton pairs of mass in the $M_Z$ region. As pointed out in \cite{Gaunt:2010pi}
$Z(\gamma^*)+jets$ production, with one of the leptons undetected, can also mimic
$W+nj$ processes.

$Z+nj$ production probes initial state parton combinations which are different from those
probed in $W^\pm W^\pm$ processes. The latter, at lowest order,
are always initiated by four--fermion states,
mainly $u\bar{d}u\bar{d}$. The former, on the contrary, typically have at least
two gluons in the initial state since the largest component \cite{Maina:2009vx,Maina:2009sj}
involves a two jet process which is dominated by gluon--gluon scattering.

For comparison we also present the predictions for $\gamma+3j$ production,
the reaction from which the most recent and precise estimates of $\sigma_{eff}$
have been extracted at the Tevatron.
This measurement will undoubtedly be performed again at the LHC
\cite{Bechtel:2009zz}.

NLO QCD corrections are or will soon be available for all SPI 
processes leading to an electroweak
vector boson in association with up to four jets
\cite{Campbell:2007ev,Campbell:2002tg,Campbell:2003hd,
Ellis:2009zw,KeithEllis:2009bu,Berger:2008sz,Berger:2009ep,Berger:2010vm}.
The Drell-Yan cross section is known at NNLO \cite{Hamberg:1990np}.
Measurements at the Tevatron show good agreement
between NLO calculations and data \cite{Aaltonen:2007ip,Aaltonen:2007cp}. These new developments
open the possibility of validating the predictions using events with large visible
energy, where the MPI
contribution is small, and then using them for a direct measurement of the MPI
cross section at smaller total invariant masses in parallel with more data driven
analysis similar to those of CDF and D0.

In the following we compare the results obtained with the
GS09 dPDF with those obtained with two instances
of fully factorized sPDF: MSTW2008LO \cite{Martin:2009iq} and CTEQ6L1 \cite{Pumplin:2002vw}.
Hence we can estimate, even in the absence of a proper dPDF set based on CTEQ6,
the dependence of MPI predictions on the choice of PDF, a study that to our knowledge
has not been performed before.

We have considered three center of mass energies for the LHC: $\sqrt{s}$ = 7 TeV,
$\sqrt{s}$ = 10 TeV and $\sqrt{s}$ = 14 TeV. This allows us to study the properties
of MPI processes while the relevant range of momentum fractions for the dPDF shifts to
smaller values as the energy increases.

Given the strong similarities between the production mechanism of
$Z+jets$, $W+jets$ and $\gamma+jets$ we expect
that the conclusions reached in the present paper for $Z+jets$ production concerning the ratio
of the MPI to the SPI contribution, the effect of correlations in MPI and the
dependence on the PDF choice will be applicable also to $W+jets$ and $\gamma+jets$
production.

We will confine ourselves to Double Parton Interactions and neglect Triple and
Higher Order Parton Interactions. Triple Parton Interactions are expected to be
significantly less abundant than Double ones, even though it has been argued that
they could be indeed detected at the LHC \cite{Maina:2009vx,Maina:2009sj}.

In \sect{sec:calc} the main features
of the calculation are discussed.
Then we present our results in \sect{sec:res}.
Finally we summarize the main points of our discussion.

%%%%%%%%%%%%%%%%%%%%%%%%%%%%%%%%%%%%%%%%%%%%%%%%%%%%%%%%%%%%%%%%%%%%%%%%%%%%%%

\section{Calculation}
\label{sec:calc}

The  MPI processes which contribute at leading order to $Z+n$--jets through Double
Parton Interactions are those in which an event producing $k$ jets is superimposed
to an event producing a $Z$--boson and $(n-k)$ jets, $k=2,\dots,n$.

At the Tevatron, CDF \cite{Abe:1997bp,Abe:1997xk}
has measured $\sigma_{eff}=14.5\pm 1.7^{+1.7}_{-2.3}$ mb, a value
confirmed by D0 which quotes
$\sigma_{eff}=15.1\pm 1.9$ mb \cite{Abazov:2009gc}.
In Ref.~\cite{Treleani:2007gi} it is argued, on the basis of the simplest two channel
eikonal model for the proton--proton cross section, that a more appropriate value at
$\sqrt{s}= 1.8$ TeV is 10 mb which translates at the LHC into  
$\sigma_{eff}^{LHC}=12$ mb. Treleani then estimates the effect of the removal by CDF
of TPI events from their sample and concludes that the CDF measurement yields
$\sigma_{eff} \approx 11$ mb at Tevatron energies.
In the following we use  $\sigma_{eff}=12.0$ mb for
all LHC center of mass energies,
with the understanding that this value is affected by an experimental uncertainty
of about 15\% and that it agrees only within 30\% with the predictions of the eikonal model.
Since $\sigma_{eff}$ appears as an overall factor
in our results it is easy to take into account a different value.

It is worth mentioning that at present there is a discrepancy between the value
of $\sigma_{eff}$ extracted by CDF and D0 and the one which is effectively
employed by Pythia
whose normalization is derived mainly from comparisons with small p$_{T}$ data
which dominate the total cross section.
The description of MPI in {\tt PYTHIA8}~\cite{Sjostrand:2007gs}
assumes that interactions can occur at different p$_{T}$ values 
independently of each other inside inelastic non--diffractive events.  
The expression for a DPI cross section becomes therefore: 
\begin{equation}
\label{MPI_eq:sigma_3}
    \sigma =  <f_{impact}>\sigma_1 \cdot \sigma_2/\sigma_{ND}/k
\end{equation}
where $\sigma_{ND}$ is the total non--diffractive cross section and $f_{impact}$
is an enhancement/depletion factor chosen event-by-event to account for correlations 
introduced by the centrality of the collision. This quantity is typically averaged during 
an entire run to calculate $<f_{impact}>$ in Eq.~\ref{MPI_eq:sigma_3}.
Typical values at the center of mass energy of 10~TeV are 1.33 for
$<f_{impact}>$ and 51.6~mb for $\sigma_{ND}$.
Comparing Eq.~\ref{MPI_eq:sigma_3} with Eq.~\ref{eq:sigma_D3} tells us that {\tt PYTHIA8}
predicts an effective $\sigma_{eff}$=$\sigma_{ND}$/$<f_{impact}>$ which
is about a factor three larger than the one actually measured at the Tevatron. I believe
that this issue deserves careful consideration and that new measurements of high p$_{T}$
MPI reactions would be quite welcome.

All samples have been generated with the following cuts:

\bea
\label{eq:cuts}
& p_{T_j} \geq 30~{\rm GeV} \, , \; \; |\eta_j| \leq 5.0 \, , 
\nonumber \\
& p_{T_\ell} \geq 20~{\rm GeV} \, ,\; \;
|\eta_{\ell}| \leq 2.5 \, , \\
& p_{T_\gamma} \geq 30~{\rm GeV} \, , \; \; |\eta_\gamma| \leq 2.5 \, , 
\nonumber \\
& \Delta R_{jj} \geq 0.1  \, ,\; \; \Delta R_{jl} \geq 0.1\, ,\; \; \Delta R_{j\gamma} \geq 0.1\nonumber 
\eea

where $j= u,\bar{u},d,\bar{d},s,\bar{s},c,\bar{c},b,\bar{b},g$ and $l=e^+,e^-,\mu^-,\mu^+$.

The $Z+4$--jets sample has been generated with \Phantom
\cite{Ballestrero:2007xq,Ballestrero:1994jn,Ballestrero:1999md}, while all
other samples
have been produced with \MadEvent \cite{Maltoni:2002qb,Alwall:2007st}.
Both programs generate equal weight events 
in the Les Houches Accord File Format \cite{Alwall:2006yp}.
All samples have been generated using CTEQ6L1 \cite{Pumplin:2005rh} 
parton distribution functions.
The QCD scale (both in $\alpha_s$ and in the parton distribution functions)
has been taken as

\be
\label{eq:LargeScale}
Q^2 =\sum_{i=1}^n p_{Ti}^2,
\ee
where $n$ is the number of final state partons, for all reactions with the exception of
$q\bar{q}\rightarrow l^+l^-$ for which the scale has been set at $Q^2 = M_Z^2$.
The scale in \eqn{eq:LargeScale} is similar, though not identical, to the scale advocated
in \rfs{Berger:2009ep,Berger:2010vm} for vector boson plus jets production at NLO.

The results shown in the following under the CTEQ heading have been obtained
combining 
at random one event from each of the reactions which together
produce the  desired final state through MPI.
When needed, we have required that each
pair of colored partons in the final state have a minimum $\Delta R$ separation.
This implies that the combined cross section does not in general correspond
to the product of the separate cross sections divided by
$\sigma_{eff}$ because the requirement of a minimum
separation for all jet pairs induces a reduction of the cross section
when additional pairs are formed in superimposing events.

The results shown under the MSTW and GS09 headings have been obtained through a
reweighting procedure by the appropriate ratio of parton distribution functions
and coupling constants. For instance, an event like $(q_i\bar{q_i} \rightarrow gl^+l^-)
\otimes (gg \rightarrow gg)$, constructed from two events generated separately
with CTEQ6 PDF,
can be transformed in a weighted event with MSTW2008 PDF multiplying its original
weight by  

\be
\label{eq:reweight}
R = \frac{F^{^{MSTW}}_i(t_1)F^{^{MSTW}}_{\bar{i}}(t_1)}{F^{^{CTEQ}}_i(t_1)F^{^{CTEQ}}_{\bar{i}}(t_1)} \times
\frac{\alpha^{^{MSTW}}_s(t_1)}{\alpha^{^{CTEQ}}_s(t_1)} \times
\frac{F^{^{MSTW}}_g(t_2)F^{^{MSTW}}_g(t_2)}{F^{^{CTEQ}}_g(t_2)F^{^{CTEQ}}_g(t_2)}  \times
\frac{\alpha^{^{MSTW}}_s(t_2)^2}{\alpha^{^{CTEQ}}_s(t_2)^2}
\ee

where $t_1,\, t_2$ are the factorization scales for $q_i\bar{q_i} \rightarrow gl^+l^-$
and $gg \rightarrow gg$ respectively. The factorization scales have been read off
from the event files.
The second and fourth factors in \eqn{eq:reweight}
take into account the different values of the strong coupling constants for the two
different sets of PDF: $\alpha_{s,LO}^{^{CTEQ}}(M_Z)=0.130$ while
$\alpha_{s,LO}^{^{MRST}}(M_Z)=\alpha_s^{^{GS09}}(M_Z)=0.139$.
The only difference for the GS09 case would be that
the correlated dPDF $F_{ij}(t_1,t_2)$ would appear instead of the uncorrelated
product $F_i(t_1)F_j(t_2)$ and so on.
The resulting events are no longer unweighted. The error on the cross section
introduced by the reweighting procedure is essentially negligible because of the
large size, about $5\times 10^5$ events, of the samples.
Reweighting can also be employed to estimate the sensitivity of our tree level results
to variations of the scale \eqn{eq:LargeScale} using a straightforward modification
of \eqn{eq:reweight}.

All results are obtained with the following values for the electroweak input parameters:
$M_Z$ = 91.188 GeV, $M_W$ = 80.40 GeV, $G_F$ = 0.116639 $\times$ 10$^{-5}$ GeV$^{-2}$. 

We work at parton level with no showering and hadronization.
Color correlations between the two scatterings have been ignored.

\section{Results}
\label{sec:res}

The total cross sections for SPI and DPI production for $Z+2$--jets, $Z+3$--jets
and $Z+4$--jets are presented in \tbn{tab:Z2j}, \tbn{tab:Z3j} and \tbn{tab:Z4j}
respectively. In all cases the cuts in \eqn{eq:cuts} have been imposed.
The analysis has been repeated requiring a larger separation between jets;
the results for $\Delta R_{jj}=0.5$ and for $\Delta R_{jj}=0.7$ are also shown.
In our estimates below we have only taken into account the muon
decay of the $Z$ boson. The $Z\ra e^+e^-$ channel gives the same result.
The possibility of detecting high $p_T$ taus has been
extensively studied in connection with the discovery of a light Higgs in Vector
Boson Fusion in the
$\tau^+\tau^-$ channel with extremely encouraging results \cite{Leney:2008di}.
Therefore, the tau decay of the $Z$ can be expected to contribute significantly to
MPI studies.

\TABLE{
%\begin{table}[thb]
\label{tab:Z2j}
%\vspace{0.15in}
%\begin{center}
\begin{tabular}{|c|c|c|c|c|c|c|c|c|c|}
\hline
 & \multicolumn{3}{|c|}{14 TeV} & \multicolumn{3}{|c|}{10 TeV}  & \multicolumn{3}{|c|}{7 TeV} \\
\hline
 $Z+2j$ &  CTEQ  & MSTW & GS09   &  CTEQ  & MSTW & GS09   &  CTEQ  & MSTW & GS09 \\
\hline
\multicolumn{10}{|c|}{$\Delta R_{jj}=0.1$} \\
\hline
SPI & 56.71 & 65.35 &    & 33.11  & 37.97  &   &  17.97   & 20.48 & \\
\hline
DPI & 11.27  & 14.37 & 15.50  & 4.80  & 6.35  & 6.68 &  1.88 & 2.61 & 2.66\\
\hline
\multicolumn{10}{|c|}{$\Delta R_{jj}=0.5$}\\ 
\hline
SPI & 52.65 & 60.70 &    & 30.63  & 35.15  &   &  16.56   & 18.88 & \\
\hline
DPI & 11.27  & 14.37 & 15.50  & 4.80  & 6.35  & 6.68 &  1.88 & 2.61 & 2.66\\
\hline
\multicolumn{10}{|c|}{$\Delta R_{jj}=0.7$} \\
\hline
SPI & 51.53 & 59.41 &    & 29.95  & 34.38  &   &  16.17   & 18.45 & \\
\hline
DPI & 11.27  & 14.37 & 15.50  & 4.80  & 6.35  & 6.68 &  1.88 & 2.61 & 2.66\\
\hline
\end{tabular}
%\end{center}
\caption{$Z+2$--jets, $Z\rightarrow\mu^+\mu^-$ cross sections in pb. Cuts as in \eqn{eq:cuts} with
increasing angular separation between jets, $\Delta R_{jj}$.
}
%\end{table}
}

\TABLE{
%\begin{table}[thb]
\label{tab:Z3j}
%\vspace{0.15in}
%\begin{center}
\begin{tabular}{|c|c|c|c|c|c|c|c|c|c|}
\hline
 & \multicolumn{3}{|c|}{14 TeV} & \multicolumn{3}{|c|}{10 TeV}  & \multicolumn{3}{|c|}{7 TeV} \\
\hline
$Z+3j$  &  CTEQ  & MSTW & GS09   &  CTEQ  & MSTW & GS09   &  CTEQ  & MSTW & GS09 \\
\hline
\multicolumn{10}{|c|}{$\Delta R_{jj}=0.1$} \\
\hline
SPI & 21.62 & 26.25 &    & 11.75  & 14.18  &   &  5.77   & 6.91 & \\
\hline
DPI & 2.93  & 4.06 & 4.20  & 1.10  & 1.61  & 1.60 &  0.37 & 0.58 & 0.55\\
\hline
\multicolumn{10}{|c|}{$\Delta R_{jj}=0.5$}\\ 
\hline
SPI & 15.71 & 19.10 &    & 8.46  & 10.23  &   &  4.11   & 4.93 & \\
\hline
DPI & 2.70  & 3.75 & 3.88  & 1.02  & 1.49  & 1.48 &  0.34 & 0.54 & 0.51\\
\hline
\multicolumn{10}{|c|}{$\Delta R_{jj}=0.7$} \\
\hline
SPI & 14.13 & 17.18 &    & 7.59  & 9.18  &   &  3.67   & 4.41 & \\
\hline
DPI & 2.59  & 3.60 & 3.73  & 0.97  & 1.43  & 1.42 &  0.33 & 0.52 & 0.49\\
\hline
\end{tabular}
%\end{center}
\caption{$Z+3$--jets, $Z\rightarrow\mu^+\mu^-$ cross sections in pb. Cuts as in \eqn{eq:cuts} with
increasing angular separation between jets, $\Delta R_{jj}$.
}
%\end{table}
}

The total cross sections for SPI and DPI production for $\gamma+3$--jets
are shown in \tbn{tab:gamma3j} with increasing jet--jet separation. 
It should be mentioned however that at the LHC
trigger thresholds for single photons are foreseen
to be much higher than those for double leptons \cite{Bayatian:2006zz,Ball:2007zza,Aad:2009wy}.
While pair of leptons are expected to be triggered on for transverse momenta of about
15 GeV, single photons will be detected only when their transverse momenta 
is larger than about 80 GeV at the design energy of 14 TeV. At lower energies
and instantaneous luminosities the threshold could be smaller.
Even at design luminosity and center of mass energy a lower threshold could be allowed
with some pre--scaling.
Since MPI processes are known to decrease sharply with increasing transverse momenta,
we present in \tbn{tab:gamma3jpT80}
the predictions for $p_{T_\gamma} \geq 80~{\rm GeV}$ while the results in \tbn{tab:gamma3j}
are mainly intended for low luminosity data taking.

\TABLE{
%\begin{table}[thb]
\label{tab:Z4j}
%\vspace{0.15in}
%\begin{center}
\begin{tabular}{|c|c|c|c|c|c|c|c|c|c|}
\hline
 & \multicolumn{3}{|c|}{14 TeV} & \multicolumn{3}{|c|}{10 TeV}  & \multicolumn{3}{|c|}{7 TeV} \\
\hline
 $Z+4j$ &  CTEQ  & MSTW & GS09   &  CTEQ  & MSTW & GS09   &  CTEQ  & MSTW & GS09 \\
\hline
\multicolumn{10}{|c|}{$\Delta R_{jj}=0.1$} \\
\hline
SPI & 8.80 & 11.16 &    & 4.23  & 5.33  &   &  1.80   & 2.25 & \\
\hline
DPI & 1.21  & 1.92 & 1.82  & 0.41  & 0.71  & 0.66 &  0.12 & 0.23 & 0.20\\
\hline
\multicolumn{10}{|c|}{$\Delta R_{jj}=0.5$}\\ 
\hline
SPI & 4.26 & 5.41 &    & 2.00  & 2.53  &   &  0.83   & 1.04 & \\
\hline
DPI & 0.96  & 1.53 & 1.50  & 0.33  & 0.56  & 0.52 &  0.10 & 0.18 & 0.16\\
\hline
\multicolumn{10}{|c|}{$\Delta R_{jj}=0.7$} \\
\hline
SPI & 3.34 & 4.24 &    & 1.56  & 1.97  &   &  0.64   & 0.80 & \\
\hline
DPI & 0.87  & 1.39 & 1.35  & 0.29  & 0.51  & 0.47 &  0.09 & 0.16 & 0.14\\
\hline
\end{tabular}
%\end{center}
\caption{$Z+4$--jets, $Z\rightarrow\mu^+\mu^-$ cross sections in pb. Cuts as in \eqn{eq:cuts} with
increasing angular separation between jets, $\Delta R_{jj}$.
}
%\end{table}
}

The Single Particle Interaction MSTW results are larger than those obtained with the
CTEQ PDF by an amount which varies between 15\% for $Z+2j$ to 27\% for $ Z+4j$, increasing
as expected with the power of $\alpha_{s}$ in the amplitude.
The Double Particle Interaction MSTW results are larger than those obtained with the
CTEQ PDF by an amount which varies between 30\% and 90\%. The larger shift is due to the
smaller scales for the two individual scatterings compared to a single interaction event
with the same final state particles.
The predictions for the GS09 correlated dPDF are larger than those with
MSTW uncorrelated ones for $\sqrt{s} = 14~{\rm TeV}$ and $\sqrt{s} = 10~{\rm TeV}$
while they are smaller for $\sqrt{s} = 7~{\rm TeV}$. The difference is at most of 15\%.
Taking into account the errors in the measurement of $\sigma_{eff}$ we conclude that
the uncertainties due to the choice of PDF and to correlation effects
are reasonably under control.

These variations should be compared with the uncertainty due to scale variation in
PDF and in the strong coupling constant. In order to estimate the latter we have reweighted
our samples changing the scale in \eqn{eq:LargeScale} by a factor of two in either direction
for two limiting cases, namely $Z+2j$ production at $\sqrt{s} = 7~{\rm TeV}$ and
$Z+4j$ production at $\sqrt{s} = 14~{\rm TeV}$. In both instances we have used
MSTW PDF and $\Delta R_{jj}=0.5$. 
For $Z+2j$ production at $\sqrt{s} = 7~{\rm TeV}$ the cross section changes by
+14\%/-13\% when the scale is halved/doubled;
in the case of $Z+4j$ production at $\sqrt{s} = 14~{\rm TeV}$ the corresponding shifts
are +57\%/-29\%. The processes we are interested in therefore are not overly sensitive
to scale variations. The corresponding uncertainty is of the same order than that
related to PDF choice.

The effects of higher order corrections are more difficult to estimate since no NLO
calculation for MPI processes is available. QCD one loop calculations are available
for vector boson production with up to four jets
\cite{Campbell:2007ev,Campbell:2002tg,Campbell:2003hd,
Ellis:2009zw,KeithEllis:2009bu,Berger:2008sz,Berger:2009ep,Berger:2010vm} and are typically
of order 10\% with the exception of Drell--Yan inclusive production
\cite{Anastasiou:2003ds} where they are of the order of 50\%.
NLO corrections for the inclusive jet cross section at the LHC have been presented in
\rf{Campbell:2006wx}. For small transverse momenta, as the ones we are interested in this paper,
they are of the order of 10\%.

\TABLE{
%\begin{table}[thb]
\label{tab:gamma3j}
%\vspace{0.15in}
%\begin{center}
\begin{tabular}{|c|c|c|c|c|c|c|c|c|c|}
\hline
& \multicolumn{3}{|c|}{14 TeV} & \multicolumn{3}{|c|}{10 TeV}  & \multicolumn{3}{|c|}{7 TeV} \\
\hline
 $\gamma+3j$ &  CTEQ  & MSTW & GS09   &  CTEQ  & MSTW & GS09   &  CTEQ  & MSTW & GS09 \\
\hline
\multicolumn{10}{|c|}{$\Delta R_{jj}=0.1$} \\
\hline
SPI & 5921.7 & 7341.4 &     & 3484.2  & 4302.5  &   &  1884.9   & 2317.1 & \\
\hline
DPI & 436.9  & 612.7 & 663.2  & 176.4  & 262.4  & 273.8 &  64.4 & 103.4 & 103.0\\
\hline
\multicolumn{10}{|c|}{$\Delta R_{jj}=0.5$}\\ 
\hline
SPI &  4516.7 & 5610.2 &    & 2637.2  & 3263.8  &   &  1415.8   & 1744.6 & \\
\hline
DPI & 422.2  & 593.7 & 642.2 & 170.3 & 254.1  & 264.9 &  62.0 & 100.0 & 99.4\\
\hline
\multicolumn{10}{|c|}{$\Delta R_{jj}=0.7$} \\
\hline
SPI & 4137.4 & 5142.9 &    & 2411.1  & 2986.4  &   &  1290.3   & 1591.3 & \\
\hline
DPI & 407.5  & 574.6 & 621.0 & 164.1  & 245.7  & 255.9 &  59.7 & 96.6 & 96.0\\
\hline
\end{tabular}
%\end{center}
\caption{$\gamma+3$--jets cross sections in pb. Cuts as in \eqn{eq:cuts} with
increasing angular separation between jets, $\Delta R_{jj}$.
}
%\end{table}
}

\TABLE{
%\begin{table}[thb]
\label{tab:gamma3jpT80}
%\vspace{0.15in}
%\begin{center}
\begin{tabular}{|c|c|c|c|c|c|c|c|c|c|}
\hline
 & \multicolumn{3}{|c|}{14 TeV} & \multicolumn{3}{|c|}{10 TeV}  & \multicolumn{3}{|c|}{7 TeV} \\
\hline
 $\gamma+3j$ &  CTEQ  & MSTW & GS09   &  CTEQ  & MSTW & GS09   &  CTEQ  & MSTW & GS09 \\
\hline
\multicolumn{10}{|c|}{$\Delta R_{jj}=0.1$} \\
\hline
SPI & 944.8 & 1142.4 &     & 524.0  & 629.9  &   &  256.8   & 306.6 & \\
\hline
DPI & 18.4  & 29.0 & 29.0  & 6.84  & 11.88  & 11.22 &  2.17 & 4.24 & 3.72\\
\hline
\multicolumn{10}{|c|}{$\Delta R_{jj}=0.5$}\\ 
\hline
\hline
SPI &  671.5 & 813.0 &    & 368.30  & 443.38  &   &  177.4   & 212.0 & \\
\hline
DPI & 17.7  & 28.1 & 28.1 & 6.59 & 11.49  & 10.85 &  2.09 & 4.09 & 3.58\\
\hline
\multicolumn{10}{|c|}{$\Delta R_{jj}=0.7$} \\
\hline
SPI & 599.7 & 726.4 &    & 328.18  & 395.29  &   &  156.6   & 187.2 & \\
\hline
DPI & 17.1  & 27.2 & 27.2 & 6.34  & 11.10  & 10.47 &  2.00 & 3.94 & 3.45\\
\hline
\end{tabular}
%\end{center}
\caption{$\gamma+3$--jets cross sections in pb. Cuts as in \eqn{eq:cuts} and
$p_{T_\gamma} \geq 80~{\rm GeV}$, with
increasing angular separation between jets, $\Delta R_{jj}$.
}
%\end{table}
}

The ratio between the MPI and SPI cross sections increases with the collider energy,
that is with decreasing average momentum fractions carried by the incoming partons.
It also increases with the $\Delta R_{jj}$ separation because of the absence of correlations
between the final state partons originating in the independent scatterings which compose
MPI events.
For $Z+nj$ processes and taking $\Delta R_{jj}=0.5$ as an example, the ratio 
is of the order of 10\% for $\sqrt{s} = 7~{\rm TeV}$ and grows to about 25\%
at $\sqrt{s} = 14~{\rm TeV}$.
The results for $\gamma+3$--jets show a similar behaviour with somewhat smaller
fractions of MPI events to SPI ones which however
depend drastically on the $p_{T_\gamma}$
cut. For $p_{T_\gamma} \geq 30~{\rm GeV}$ they range between 5 and 10\% while
for $p_{T_\gamma} \geq 80~{\rm GeV}$ they are at the percent level.

If we consider the MPI processes as our signal and the SPI ones
as the corresponding background,
we can estimate the prospect of measuring MPI in a given final state from the standard
$S/\sqrt{B}$ significance.
Using for $S$ the result obtained with GS09 PDF and for $B$ the result for the MSTW set
and assuming a luminosity of one inverse femtobarn at 7 TeV, the significancies
extracted from \tbns{tab:Z2j}{tab:Z4j}, in the $Z\rightarrow\mu^+\mu^-$ channel alone,
are 19/7/5 for $Z$+2/3/4 jets with $\Delta R_{jj}=0.7$.
The corresponding number of expected MPI events are 2600/500/140.
Therefore it appears quite feasible to measure the MPI contribution to  
$Z$+2/3/4 jets already in the first phase of the LHC.

The significance of $\gamma+3$--jets depends on the trigger strategies.
If the threshold for single photon detection can be brought in the $30~{\rm GeV}$
range then the much larger production rate, about ten times that of $Z(\mu\mu)+2j$,
provides the best opportunity for an early measurement of MPI at the LHC.
If, on the contrary, the photon trigger cannot substantially deviate from about
$80~{\rm GeV}$, $Z+2j$ production looks more promising than the $\gamma+3$--jets channel
whose significance becomes similar to that of $Z+3j$.
Anyway, in order to go beyond measuring $\sigma_{eff}$ and start to
extract the double parton distribution functions from the data, 
one should
measure the MPI fraction of as many channels as possible, exploiting the fact that
different reactions are initiated by different combinations of partons.

\begin{figure}[tb]
\centering
\subfigure{	 
\hspace*{-2.3cm} 
\includegraphics*[width=8.3cm,height=6.2cm]{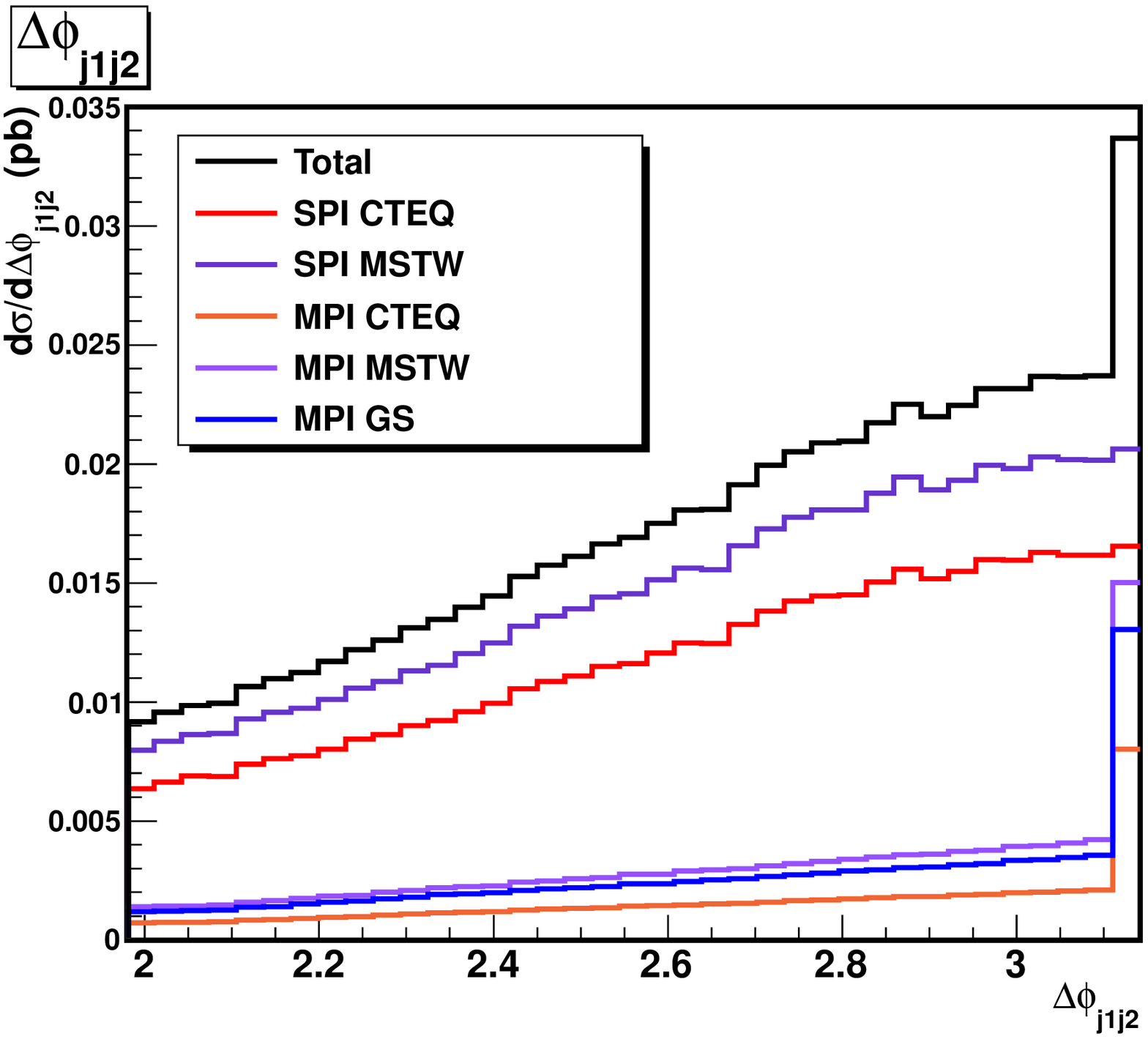}
%\hspace*{-0.7cm}
\includegraphics*[width=8.3cm,height=6.2cm]{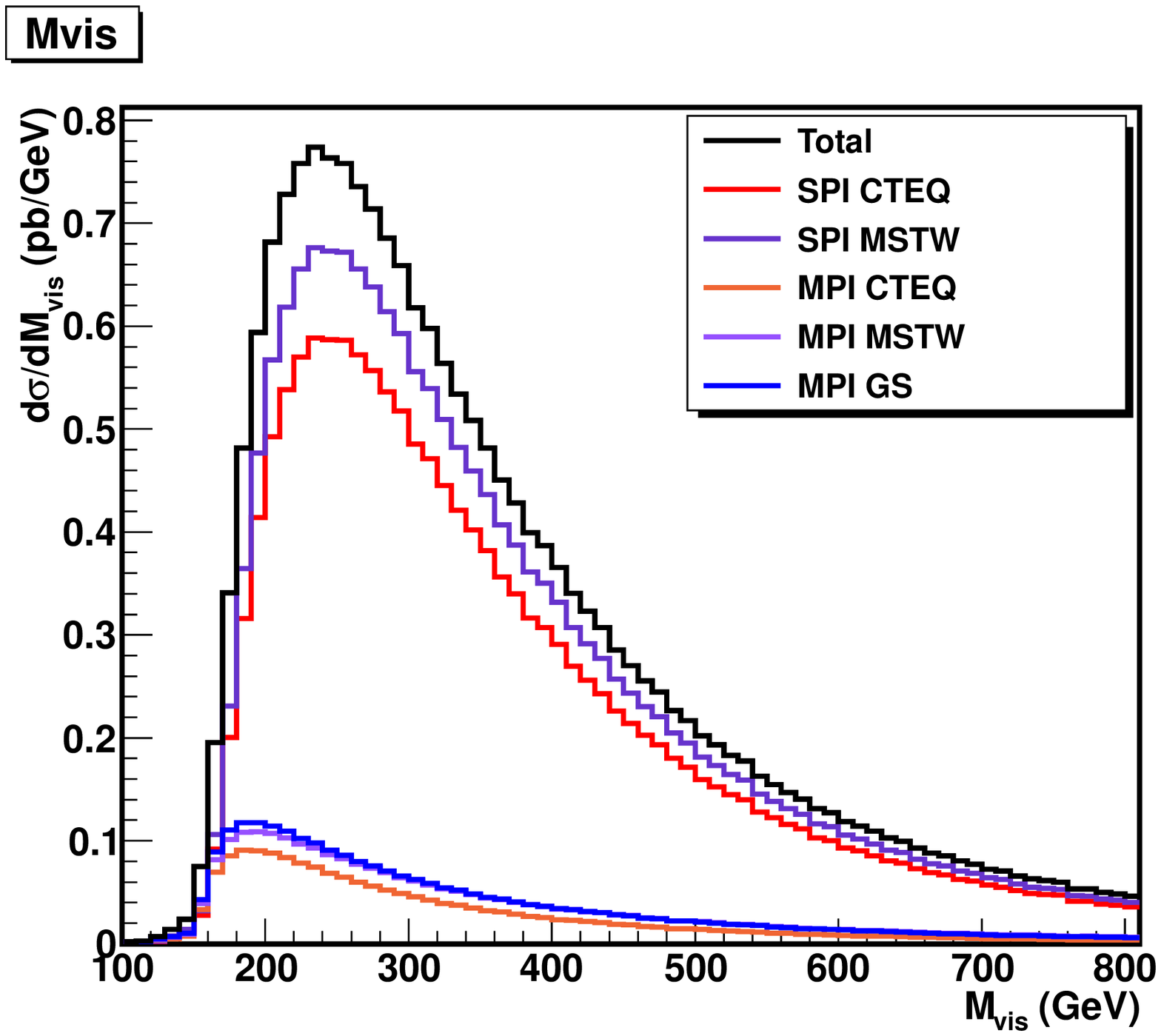}
\hspace*{-3cm}
}
\caption{On the left: distribution of the angular separation 
in the transverse plane between the two highest $p_{T}$ jets  in $Z+4j$ events.
On the right: distribution of the total visible mass,
$(\sum_{i=1}^n p_{i})^2$, in $Z+2j$ events. For both plots $\sqrt{s} = 7~{\rm TeV}$,
$\Delta R_{jj}=0.7$.}
\label{fig:Zjets_dist}
\end{figure}

The contribution to the MPI $Z+n$--jets cross section
due to two jet production in association to $Z+(n-2)$--jets processes
is in all instances the largest one,
therefore, even with more than two jets in the final state, the majority of MPI
events are expected to contain a pair of jets which are back to back in the transverse plane.
This is confirmed by the left hand side of \fig{fig:Zjets_dist}
which displays the distribution of the angular separation $\Delta\phi$
between the two highest $p_{T}$ jets in $Z+4j$ events
at $\sqrt{s} = 7~{\rm TeV}$ and $\Delta R_{jj}=0.7$.

%\hfill
%\eject

The right hand side of \fig{fig:Zjets_dist} presents the 
total visible mass distribution in $Z+2j$ production with
the same energy and angular separation. It clearly shows that MPI events are produced with
a smaller center of mass energy than SPI ones.
Whether or not these different kinematical distribution can be exploited to further
increase the MPI fraction in the event sample depends on the behaviour of the additional radiation
produced in association with the hard scattering(s) which is bound to distort both
the total visible mass and the relative orientation of jet pairs.
A dependable estimate of these effects require to pass the hard events to a
showering Monte Carlo, keeping in mind the normalization uncertainties mentioned
in \sect{sec:calc}.

\begin{figure}[bth]
\centering
\includegraphics*[width=8.3cm]{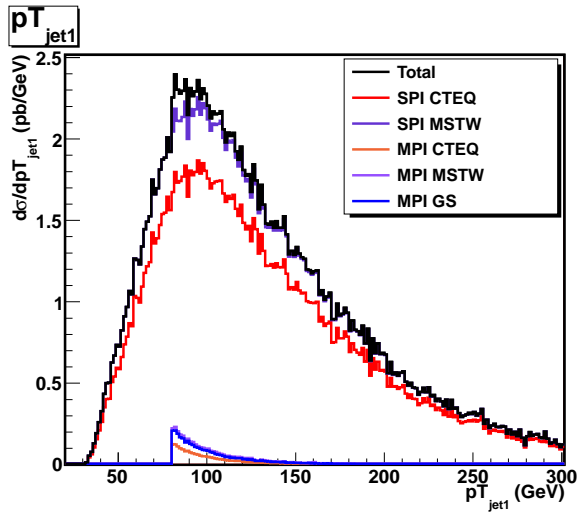}
\caption{Transverse momentum distribution of the hardest jet in
$\gamma+3j$ events.  $\sqrt{s} = 7~{\rm TeV}$,
$\Delta R_{jj}=0.7$ and $p_{T_\gamma} \geq 80~{\rm GeV}$.}
\label{fig:gammajets_dist}
\end{figure}

The only MPI mechanism contributing at tree
level to $\gamma+3$--jets is the production of two jets in one scattering and of
a photon and a jet in the other. 
Therefore, when the photon threshold is large, a jet of comparable transverse momentum
is also present. This feature could reasonably be expected to provide an additional tool
to significantly reduce the SPI contribution. Unfortunately, as shown in \fig{fig:gammajets_dist},
only a modest reduction can be achieved in this way since the $p_{T}$ spectrum
of the highest transverse momentum jet is quite hard in SPI events.

%%%%%%%%%%%%%%%%%%%%%%%%%%%%%%%%%%%%%%%%%%%%%%%%%%%%%%%%%%%%%%%%%%%%%%%%%%%%%%%%

\section{Conclusions}
\label{sec:conclusions}

In this paper we have estimated the contribution of
Multiple Parton Interactions to $Z+2/3/4$--jets and $\gamma+3$--jets production,
comparing the traditional factorized double parton distribution functions,
using both MSTW2008LO  and CTEQ6L1 PDF, and the
new correlated set by Gaunt and Stirling.

The predictions for the GS09 correlated dPDF differ by at most 15\% from those with
MSTW uncorrelated distribution functions.
The uncertainty due to the choice of PDF is in the 30 to 90\% range.

It appears quite feasible to measure the MPI contribution to  
$Z$+2/3/4 jets already in the first phase of the LHC with a total
luminosity of one inverse femtobarn at 7 TeV. If as expected the trigger threshold
for single photons is around $80~{\rm GeV}$, the $Z+2$--jets process
may well turn out to be more easily reachable than the $\gamma+3$--jets channel.
It is worth recalling that the results presented here are expected to be valid also
for $W+2/3/4$--jets with a larger cross section.
The possibility of measuring the MPI fraction in several channels could allow
to extract double parton distribution functions from the data.

\newpage

\section *{Acknowledgments}
We wish to
express our gratitude to Jonathan Gaunt and James Stirling for providing the grids and
interpolating routines for the PDF set of Ref.~\cite{Gaunt:2009re}.
This work has been supported by MIUR under contract 2008H8F9RA$\_$002 and by the
European Community's Marie-Curie Research 
Training Network under contract MRTN-CT-2006-035505 `Tools and Precision
Calculations for Physics Discoveries at Colliders'

\bibliography{MPI_Zjets_bibtex}
\end{document}